\documentclass[pra,reprint,amsmath,amssymb]{revtex4-1}
\usepackage{graphicx,color,amsmath,amssymb} 
\usepackage[colorlinks=true,urlcolor=blue, linkcolor=blue]{hyperref}

\def\({\left(}
\def\){\right)}

\def\av#1{\left\langle #1\right\rangle}

\def\eq#1{Eq.~(\ref{eq:#1})}

\def\eqs#1#2{Eqs.~(\ref{eq:#1}-\ref{eq:#2})} 
\def\Eqs#1#2{Equations~(\ref{eq:#1}-\ref{eq:#2})} 
\def\fig#1{Fig.~\ref{fig:#1}}
\def\Fig#1{Figure~\ref{fig:#1}}
\begin{document}

\title{Large-scale multipartite entanglement in the quantum optical frequency comb of a depleted-pump optical parametric oscillator}
\author{Reihaneh Shahrokhshahi}
\email{rs2vw@virginia.edu}
\author{Olivier Pfister}
\email{opfister@virginia.edu}
\affiliation{Department of Physics, University of Virginia, 382 McCormick Rd., Charlottesville, VA 22903, USA}

\begin{abstract}
We show theoretically that multipartite entanglement is generated on a massive scale in the spectrum, or optical frequency comb, of a single optical parametric oscillator (OPO) emitting well above threshold. In this system, the quantum dynamics of the strongly depleted pump field are responsible for the onset of the entanglement by correlating the two-mode squeezed, bipartite-entangled pairs of OPO signal fields. (Such pairs are independent of one another in the undepleted, classical pump approximation.) We verify the multipartite nature of the entanglement by evaluating the van Loock-Furusawa criterion for a particular set of entanglement witnesses deduced from physical considerations. 
\end{abstract}

\maketitle

\onecolumngrid

\section{Introduction}

The generation of massively entangled states is of great importance for quantum information. For quantum communication, good examples are multiparty quantum teleportation \cite{Yonezawa2004} and quantum secret sharing \cite{Lance2004}. For measurement-based quantum computing, cluster states \cite{Briegel2001,Zhang2006} are known to enable one-way quantum computing \cite{Raussendorf2001,Menicucci2006}. Constructing large-scale quantum registers and processors is therefore one of the prime objectives of experimental quantum information, along with the suppression or alleviation of decoherence. 

In most cases, the approach to scaling up the size of quantum registers or processors is a ``bottom-up'' one, in which individual Qbits (following Mermin's more harmonious spelling \cite{Mermin2003}) are put together to form, say, an entangled quantum register \cite{Ladd2010}. Now, there are, indeed, extremely few examples of ``top-down'' approaches to multipartite entanglement, in which a single physical system enables intrinsic generation of multipartite entanglement over a large scale. To the best of our knowledge, there are but two such systems. The first one is the individually trapped atoms in an optical lattice initially loaded with a Bose-Einstein condensate subsequently undergoing a Mott insulator transition \cite{Greiner2002}. The second one is the ensemble of entangled quantum modes of light, a.k.a.\ ``Qmodes,'' defined by the resonant frequencies---or quantum optical frequency comb (QOFC)---of an optical parametric oscillator (OPO), in which the QOFC is entangled by the OPO's nonlinear crystal \cite{Menicucci2008,Flammia2009}. Recently, the simultaneous generation of 15 identical quadripartite ``square'' cluster states was realized experimentally over 60 Qmodes of a single OPO \cite{Pysher2011}. 

The QOFC entanglement experiments mentioned above necessitate an exquisitely sophisticated OPO \cite{Pysher2011}, operated below threshold \cite{Midgley2010a}, and in which two or three different nonlinear interactions must be phasematched \cite{Pooser2005,Pysher2010}.

In this paper, we present the theoretical discovery of massive multipartite entanglement generation in a much simpler system and in a completely different regime. The system is but a standard OPO, in which only one nonlinear interaction is phasematched. In addition, the OPO must be operated well above threshold. It is somewhat surprising that such a simple, well-known system might lend itself to the generation of such an exotic quantum state as a massively multipartite one. In particular, we emphasize that the entangling interaction is only pairwise. It is the fact that all entangled pairs are derived from the same, strongly depleted pump field that generates the multipartite entanglement by way of a \emph{bona fide} 3-field Hamiltonian. This is therefore a fundamentally different situation from that of the below-threshold OPO in which pairwise interactions are chained and all their pump fields are undepleted, yielding quadratic nonlinear interactions   \cite{Menicucci2008} in lieu of cubic ones.

This paper is organized as follows. In Section 2, we introduce the system Hamiltonian, distinguishing between the depleted and undepleted pump cases. In Section 3, we solve the equations of motion for the system by employing a linearization procedure. We are certainly aware that more sophisticated treatments exist \cite{Vyas1995,Dechoum2004,Bradley2005,Midgley2010,Midgley2010a} and may indeed be interesting to use in order to explore this system further. In particular, it is worth mentioning the new physics of noncritical squeezing generation---that is, squeezing independent of the system parameters such as pump amplitude---in the transverse spatial modes of an OPO, which was recently predicted via the phenomena of spontaneous symmetry breaking \cite{Navarrete2008,Navarrete2010} and pump clamping \cite{Navarrete2009}, the latter having already been observed in the laboratory \cite{Chalopin2010}. Also, in this novel regime, the well-known, laser-like (and usually slow) phase diffusion process of an OPO \cite{Reid1989a,Courtois1991} becomes entwined with the squeezed variables and affects detection \cite{Navarrete2010}, which isn't usually the case in a critically squeezing two-mode OPO \cite{Reid1989a,Courtois1991}. As the present paper doesn't pertain to noncritical squeezing, we have set aside this issue of phase diffusion for further studies, under the hypothesis that its effect may be similar to that in the usual critical squeezing situation. We therefore focus here on the nontrivial new results obtained from the simple approach adopted here. In Section 4, we use the multipartite inseparability criterion derived by van Loock and Furusawa \cite{vanLoock2003a} to establish the existence of multipartite entanglement in the optical frequency comb of a single OPO. We then conclude. 

\section{The quantum optical frequency comb of a single OPO above threshold}

\subsection{Hamiltonian of the system}

\begin{figure}[htbp]
\begin{center}
\includegraphics[width=0.5\columnwidth]{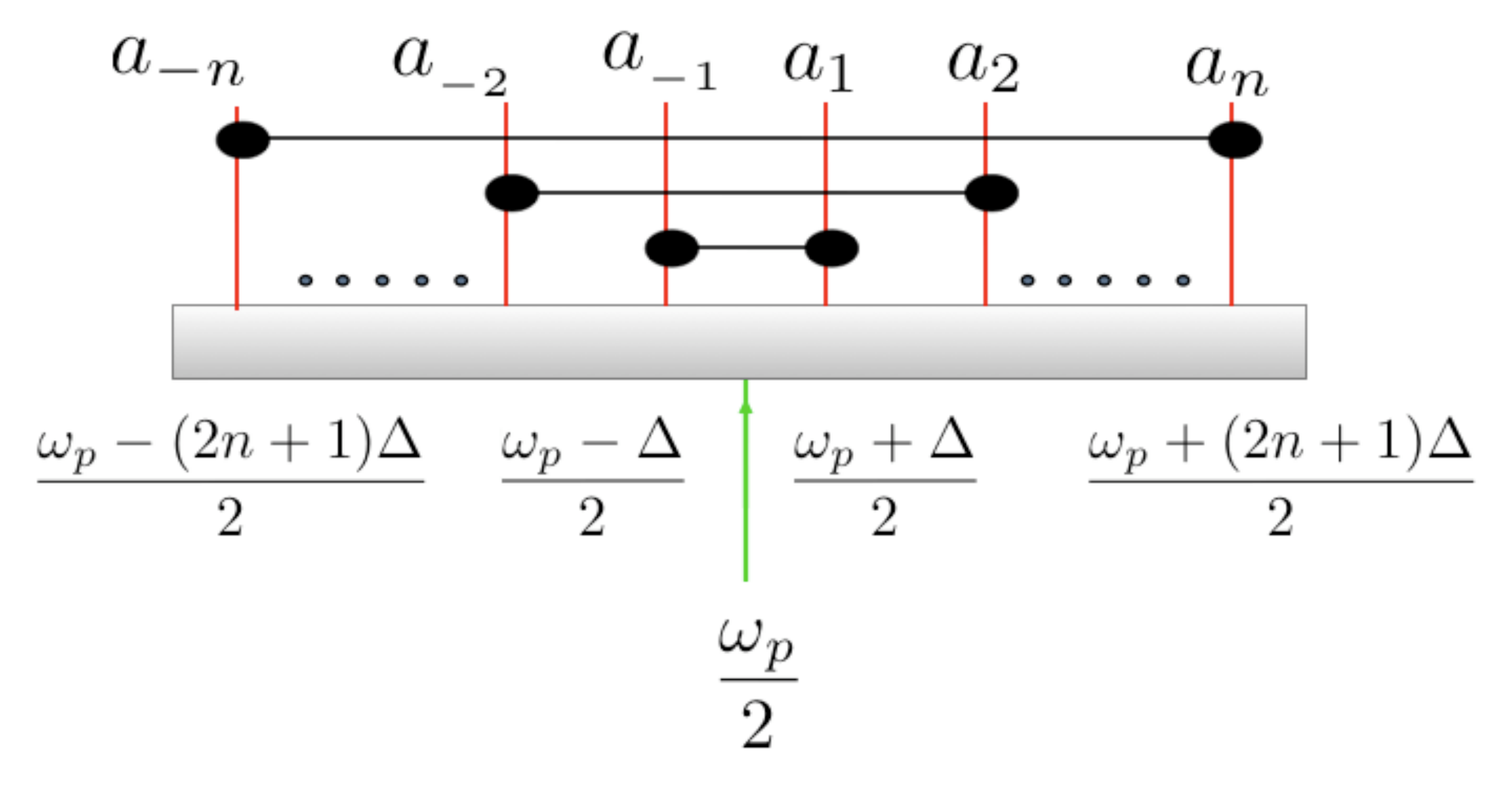}
\caption{Optical frequency comb defined by the resonant modes of the OPO cavity, spaced by free spectral range $\Delta$. The green arrow symbolizes the pump field, placed at half its frequency for clarity.}
\label{fig:comb}
\end{center}
\end{figure}

We consider the simplest possible case of an OPO with a single, nondegenerate nonlinear interaction. In this case the interaction-picture Hamiltonian is
\begin{equation}
H_{int}=2i\hbar\chi \beta\sum_{i=1}^{n} (a_{i}^{\dag} a_{-i}^{\dag}-a_{i} a_{-i}),\label{eq:hundep}
\end{equation}
where $\beta$ is the classical (real) and constant (undepleted) pump field (in practice a stable, narrow-linewidth, continuous-wave laser) and $a_{\pm i}$ are the photon annihilation operators of entangled Qmodes $\pm i$, of frequencies 
\begin{equation}
\omega_{\pm i}=\frac{\omega_p}2 \pm \(i+\frac12\)\Delta,
\end{equation}
where $\omega_p=\omega_i+\omega_{-i}$ is the pump frequency (see \fig{comb}) 
%%%%%%%%%%%%%%%%%%%%%%%%%%%%%%%%%%%%%%%%%%%%%%%%%%%
%%%%%%%%%%%%%%%%%%%%%%%%%%%%%%%%%%%%%%%%%%%%%%%%%%%
and $\Delta$ is the free spectral range of the OPO cavity.

Below the OPO emission threshold, such a system is known to emit two-mode squeezed fields which demonstrate the Einstein-Podolsky-Rosen (EPR) paradox \cite{Reid1989,Ou1992}. Above the OPO threshold, the undepleted classical pump approximation can still be taken to hold and the generation of EPR states has also been shown to be possible, theoretically \cite{Reid1988} and experimentally \cite{Villar2005,Su2006,Jing2006,Keller2008}.\footnote{Note that all previous works featured the entanglement of a single Qmode pair at a time, which is not the situation described by \eq{hundep}. Indeed, \eq{hundep} predicts many independent EPR pairs. Experimentally, this requires that the OPO cavity be resonant for all Qmode EPR pairs, which can be realized either by compensating birefringence in a type-II OPO or by using a type-I OPO. Dispersion is neglected in this discussion as its effects can be neglected for the first tens to hundreds of modes.} 

However, the undepleted classical pump approximation breaks down if the external pump power is increased significantly above threshold. In that case, the pump must be treated as a quantum field in the three-wave mixing interaction
\begin{equation}
H_{int}=2i\hbar\chi\sum_{i=1}^{n} (p a_{i}^{\dag} a_{-i}^{\dag}-p^{\dag }a_{i} a_{-i}),
\label{eq:3wm}
\end{equation}
where $p$ is the annihilation operator of the pump field.
Recently, it was predicted \cite{Villar2006} and experimentally demonstrated \cite{Coelho2009} that the pump field  participates in three-way entanglement in this case. Another interesting  theoretical analysis showed that the signal fields from two OPOs pumped by the same field could become entangled \cite{Cassemiro2008}. Here, we extend this analysis to the QOFC of a \emph{single} OPO, in which a vast number of different Qmode pairs are already known to be entangled by their parametric downconversion from the pump field \cite{Pfister2004,Pysher2011}. In the undepleted pump approximation, all EPR pumps are independent. However, when one considers the OPO well above threshold, there is but a single quantum pump field, whose strong (ideally total) depletion entail strong correlations between the EPR fields, since a pump photon downconverting into one Qmode pair will necessarily \emph{not} be downconverted into any other pair. This paper posits that this situation should yield multipartite, rather than bipartite, quantum correlations and our goal is to ascertain whether they result in multipartite entanglement, which they do.

\section{The quantum optical frequency comb of a single OPO above threshold}

As mentioned before, we consider the simplest possible case of an OPO cavity with a single pump mode and a single nondegenerate interaction in its nonlinear crystal. Such a crystal implements the  Hamiltonian of \eq{3wm}. We assume a single two-mirror standing wave cavity, with one mirror of reflectivity $R^{'}_{\pm i}=1$ for all modes and the other an output coupler of $R_{\pm i}=1-T_{\pm i} < 1$.
Taking into account the vacuum modes $A_{i}^\mathrm{in}$ that enter the cavity through its output coupler, the input-output theory \cite{Collett1984,Gardiner2004} can be used to derive the equations of motion for the internal cavity modes
\begin{align}
\dot{a_{i}} & =2\chi pa_{-i}^{\dag}-k_{i}\, a_{i}+ \sqrt{2 k_{i}}\ A_{i}^\mathrm{in}\label{eq:Heis1}\\
\dot{a}_{-i}&=2\chi p a_{i}^{\dag}-k_{-i}\, a_{-i}+ \sqrt{2 k_{-i}}\ A_{-i}^\mathrm{in}\label{eq:Heis2}\\
\dot{p}&=-2\chi\sum_{i}^{n}a_{i} a_{-i}-k_{p}\, p+ \sqrt{2 k_{p}}\ p_{in}.\label{eq:Heis3}
\end{align}
Here $k_{\pm i}={T_{\pm i}}/{2\tau}$  are the loss rates of the cavity mirror for mode $i$,  $\tau$ being the cavity round trip time.
In order to solve \eqs{Heis1}{Heis3} we first rewrite field operators as centered fluctuations about their expectation value
\begin{align}
a_{i}&=\alpha_{i}+\delta a_{i}  \\
A_{i}^\mathrm{in}&=\delta A_{i}^\mathrm{in} \\
 p&=\varpi+\delta p  \\
  p_{in}&=\varpi _{in}+\delta p_{in}.
\end{align}

\subsection{Classical steady-state solutions}

The semiclassical, or mean value, equations follow directly from \eqs{Heis1}{Heis3}:
\begin{align}
\dot{\alpha_{i}}&=2\chi\varpi  \alpha_{-i}^{\ast}-k_{{i}}\, \alpha_{i}\label{eq:Heis1meanclas}\\
\dot{\alpha}_{-i}&=2\chi\varpi  \alpha_{i}^{\ast}-k_{{-i}}\, \alpha_{-i}\label{eq:Heis2meanclas}\\
\dot{\varpi }&=-2\chi\sum_{i}^{n}\alpha_{i} \alpha_{-i}-k_{p}\, \varpi + \sqrt{2 k_{p}} \ \varpi _{in}.\label{eq:Heis3meanclas}
\end{align}
Now, considering same cavity losses for all signal modes, $k_i=k_{-i}=k_a$, the stationary solutions of the two coupled equations \eqs{Heis1meanclas}{Heis2meanclas} for semi-classical mean values are
\begin{align}
2\chi\varpi  \alpha_{-i}^{{0\ast }}&=k_a\, \alpha_{i}^{0}\\
2\chi\varpi  \alpha_{i}^{0\ast }&= k_a\, \alpha_{-i}^{0}.
\end{align}
this results in $|\alpha_{i}^{0}|=|\alpha_{-i}^{0}| $ , $|\varpi ^{0}|={k_{a}}/{2\chi}$ and $ \phi_{i}^{0}+\phi_{-i}^{0}-\phi^{0}=0 $
where $\phi_{\pm i}^{0}$ and  $\phi^{0}$  are the respective phases of $\alpha_{\pm i}^{0}$ and $\varpi ^{0}$. For simplicity we take $\varpi _{in}$ real and positive therefore based on \eq{Heis3meanclas}, $\phi^{0}=0$ and  $ \phi_{i}^{0}=-\phi_{-i}^{0}$. The stationary solution for the pump field's mean value can then be written as
\begin{align}
4\chi^2\sum_{i}^{n}|\alpha_{i}^{0}|^{2}=k_{a}k_{p}(\sqrt{\sigma}-1),\label{eq:thresh}
\end{align}
where
\begin{align}
\sigma=\(\frac{2\chi}{k_{a}}\sqrt{\frac{2}{k_{p}}}\varpi _{in}\)^{2}.
\end{align}
Clearly, the right-hand side of \eq{thresh} must be positive and $\sigma=1$ defines the threshold  pump field
\begin{equation}
\varpi _{in}^\mathrm{th}= \frac{k_a}{2\chi}\sqrt\frac{k_p}2,
\end{equation}
Hence, $\sigma=({\varpi _{in}}/{\varpi _{in}^\mathrm{th}})^{2}$ is also the pump to threshold power ratio. The classical signal amplitudes are weakly set by \eq{thresh}. 

\subsection{Stability analysis} 

The stability of the steady-state solution can be determined by a linearized analysis for small perturbations:
 \begin{align}
\alpha_{i}&=\alpha^{0}_{i}+\delta \alpha_{i}  
   \label{eq:perturb1}\\
 \varpi _{i}&=\varpi ^{0}_{i}+\delta \varpi _{i}. 
   \label{eq:perturb2}
\end{align}
Substituting \eqs{perturb1}{perturb2} into \eqs{Heis1meanclas}{Heis3meanclas}, we get
\begin{align}
&\vdots \nonumber\\
\delta \dot{ \alpha_{i}}&= k_{a} \delta \alpha_{-i}^{\ast}+ 2\chi\alpha^{0}_{i} \delta \varpi -k_{a}\delta \alpha_{i}\label{eq:Heis1meanclaspert}\\
\delta \dot{ \alpha_{-i}}&= k_{a} \delta \alpha_{i}^{\ast}+ 2\chi\alpha^{0}_{-i} \delta \varpi -k_{a}\delta \alpha_{-i}\label{eq:Heis2meanclaspert}\\
& \vdots \nonumber\\
\delta \dot{\varpi }&=-2\chi\sum_{i}^{n}(\alpha_{i}^{0} \delta \alpha_{-i}+\alpha_{-i}^{0} \delta \alpha_{i})-k_{p} \delta \varpi,  \label{eq:Heis3meanclaspert}
\end{align}
where $i=1,...,n$, $n$ being the number of signal and idler pairs considered inside the cavity. 
Defining $ \delta A=\begin{pmatrix} \dots & \delta \alpha_{i}& \delta \alpha^{*}_{i} & \delta \alpha_{-i} & \delta \alpha^{*}_{-i} & \dots  & \delta \varpi  &\delta \varpi ^{*} \end{pmatrix}^T$
We can rewrite \eqs{Heis1meanclaspert}{Heis3meanclaspert} in block matrix form:
\begin{equation} %\small
\frac d{dt}\begin{pmatrix} \vdots \\ \delta \alpha_{i}\\ \delta \alpha^{*}_{i} \\ \delta \alpha_{-i} \\ \delta \alpha^{*}_{-i}\\ \vdots   \\ \delta \varpi  \\ \delta \varpi ^{*} \end{pmatrix}
= 
\begin{pmatrix} 
&\vdots &\vdots &\vdots &\vdots & &\vdots &\vdots &\\  
 \dots &-k_{a}&0&0&k_{a}& \dots &2\chi \alpha_{i}^{0}&0\\
 \dots &0&-k_{a}&k_{a}&0& \dots &0&2\chi \alpha_{i}^{0 \ast} \\ 
 \dots &0&k_{a} & -k_{a} &0& \dots &2 \chi \alpha_{-i}^{0} &0 \\
 \dots &k_{a}&0&0&-k_{a}& \dots &0&2\chi \alpha_{-i}^{0\ast} \\
&\vdots &\vdots &\vdots &\vdots & &\vdots &\vdots &\\  
 \dots &-2\chi \alpha_{-i}^{0}&0&-2\chi \alpha_{i}^{0}&0&\dots &-k_{p}&0 \\  
 \dots &0&-2\chi \alpha_{-i}^{0 \ast }&0&-2\chi \alpha_{i}^{0 \ast} &\dots &0&-k_{p}  
 \end{pmatrix} 
 \begin{pmatrix} \vdots \\ \delta \alpha_{i}\\ \delta \alpha^{*}_{i} \\ \delta \alpha_{-i} \\ \delta \alpha^{*}_{-i}\\ \vdots \\ \delta \varpi  \\ \delta \varpi ^{*} 
 \end{pmatrix}. 
 \label{eq:matrixeq}
\end{equation}
%\begin{equation} 
%\frac d{dt}\begin{pmatrix}   \delta \varpi  \\ \delta \varpi ^{*}\\ \vdots \\ \delta \alpha_{i}\\ \delta \alpha^{*}_{i} \\ \delta \alpha_{-i} \\ \delta \alpha^{*}_{-i} \\ \vdots \end{pmatrix}
%= 
%\begin{pmatrix} 
%-k_{p}&0 & \dots &-2\chi \alpha_{-i}^{0}&0&-2\chi \alpha_{i}^{0}&0&\dots  \\  
%0&-k_{p} & \dots &0&-2\chi \alpha_{-i}^{0 \ast }&0&-2\chi \alpha_{i}^{0 \ast} &\dots   \\
%\vdots &\vdots & &\vdots &\vdots &\vdots &\vdots & &\\  
%2\chi \alpha_{i}^{0}&0 & \dots &-k_{a}&0&0&k_{a}& \dots \\
%0 & 2\chi \alpha_{i}^{0 \ast} & \dots &0&-k_{a}&k_{a}&0& \dots  \\ 
%2 \chi \alpha_{-i}^{0} &0 & \dots &0&k_{a} & -k_{a} &0& \dots  \\
%0&2\chi \alpha_{-i}^{0\ast} & \dots &k_{a}&0&0&-k_{a}& \dots  \\
%\vdots &\vdots & &\vdots &\vdots &\vdots &\vdots & 
% \end{pmatrix} 
% \begin{pmatrix} \delta \varpi  \\ \delta \varpi ^{*} \\ \vdots \\ \delta \alpha_{i}\\ \delta \alpha^{*}_{i} \\ \delta \alpha_{-i} \\ \delta \alpha^{*}_{-i}\\ \vdots  
% \end{pmatrix} 
% \label{eq:matrixeq}
%\end{equation}
We derived the eigenvalues of the matrix in \eq{matrixeq} for $n=1,2,3$.  In all these cases, the eigenvalue sets have the following form:
\begin{align}
\{\lambda\}=\{
\underset{2n-1}{\underbrace{0,\dots, 0}},\underset{2n-1}{\underbrace{-2k_{a},\dots,-2k_a}},\lambda_{1},\lambda_{2},\lambda_{3},\lambda_{4}\},
\label{eq:eigenvaluevector}
\end{align}
where% 
%\begin{align}
%\lambda_{1,2}&=-\frac{k_p}{2} \pm \frac{1}{2}\sqrt{k_p \left[k_p+8k_a(1-\sqrt\sigma)\right]}\\
%\lambda_{3,4}&=-\frac{1}{2} k_a \left( 2+k \pm \sqrt{(2+k)^2-8 n \sqrt{\sigma }}\right).
%\end{align}
, posing the pump-signal loss ratio $\kappa={k_p}/{k_a}$,
\begin{align}
\lambda_{1,2}&=-\frac{1}{2} k_a \left(\kappa \pm \sqrt{\kappa \left[\kappa-8(\sqrt\sigma-1)\right]}\right)
\label{eq:stab12}\\
\lambda_{3,4}&=-\frac{1}{2} k_a \left( \kappa+2 \pm \sqrt{(\kappa+2)^2-8 n \sqrt{\sigma }}\right).
\label{eq:stab34}
\end{align}
Because of the particular symmetry of the problem---namely the block structure of the matrix in \eq{matrixeq}, we argue that it is reasonable to postulate that \eq{eigenvaluevector} is the general eigenvalue set, $\forall n$, even though a complete inductive proof is formally required. For certain initial conditions all $2(2n+1)$ eigenvalues  of the matrix in \eq{eigenvaluevector} can only be zero or negative, which ensures  the stability of the stationary solution presented in \eq{thresh}. \Eqs{stab12}{stab34} show that, as the number of times above threshold $\sigma$ increases, one can always find negative values for $\lambda_{1,\dots,4}$ by increasing the pump-signal loss ratio $\kappa$, thereby tending towards the doubly resonant OPO, which is always stable. 
 
\subsection{Quantum fluctuations}
 
 Now, we rewrite \eqs{Heis1}{Heis3} for the quantum fluctuations around these classical mean values.
 Notice $\alpha_{-i}^{\ast}=\alpha_{i}$,
\begin{align}
\dot{\delta a_{i}}&=2\chi(\delta p\alpha_{i}e^{i\phi_{i}}+\varpi  \delta a_{-i}^ {\dag} )-k_{a} \delta a_{i}+ \sqrt{2 k_{a}} \delta A_{i}^\mathrm{in}%\label{eq:qheis1}
\\
\dot{\delta a_{-i}}&=2\chi(\delta p\alpha_{i}e^{-i\phi_{i}}+\varpi  \delta a_{i}^{\dag }) -k_{a} \delta a_{-i} + \sqrt{2 k_{a}} \delta A_{-i}^\mathrm{in}\\
\delta \dot{p}&=-2\chi\sum_{i}^{n}(\alpha_{i}e^{i\phi_{i}}\delta a_{-i}+ \alpha_{i}e^{-i\phi_{i}}\delta a_{i})-k_{p}\delta p+ \sqrt{2 k_{p}} \delta p_{in}.\label{eq:qheis3}
\end{align}
 We introduce the generalized field quadrature operators as $ Q_{i}=(e^{i\phi_{i}}a_{i}^{\dag}+e^{-i\phi_{i}}a_{i})$ and $
P_{i}=i(e^{i\phi_{i}}a_{i}^{\dag}-e^{-i\phi_{i}}a_{i})$.
Then solve these coupled equations we can use the symmetry of the equations in the exchange of the two signal modes and  introduce the new variables \cite{Reynaud1987}
\begin{align}
Q_{i+}&=Q_{{i}}+Q_{{-i}}\\
Q_{i-}&=Q_{{i}}-Q_{{-i}}\\
P_{i+}&=P_{{i}}+P_{{-i}}\\
P_{i-}&=P_{{i}}-P_{{-i}}.
\end{align}
The equations of motion for these quadratures are
\begin{align}
{\delta\dot{Q}_{i+}}  & = 4\chi\alpha_{i}\delta{Q}_{p}+\sqrt{2k_{a}}\ \delta{Q}_{i+}^\mathrm{in}
\label{eq:Heiquad1}\\
{\delta\dot{Q}_{i-}}  & = -2k_{a}\delta{Q}_{i-}+\sqrt{2k_{a}}\ \delta{Q}_{i-}^\mathrm{in}\label{eq:Heiquad2}\\
{\delta\dot{Q}_{p}} & =  -2\chi\sum_{i}^{n}\alpha_{i}\delta{Q}_{i+}-k_{p}\delta{Q}_{p}+\sqrt{2k_{p}}\ \delta{Q}_p^\mathrm{in}
\label{eq:Heiquad3}\\
{\delta\dot{P}_{i+}} & =  4\chi\alpha_{i}\delta{P}_{p}-2k_{a}\delta{P}_{i+}+\sqrt{2k_{a}}\ \delta{P}_{i+}^\mathrm{in}
\label{eq:Heiquad4}\\
{\delta\dot{P}_{i-}} & =  \sqrt{2k_{a}}\ \delta{P}_{i-}^\mathrm{in}
\label{eq:Heiquad5}\\
{\delta\dot{P}_{p}} & =  -2\chi\sum_{i}^{n}\alpha_{i}\delta{P}_{i+}-k_{p}\delta{P}_{p}+\sqrt{2k_{p}}\ \delta{P}_p^\mathrm{in}.
\label{eq:Heiquad6}
\end{align}
 As seen from \eq{Heiquad2} and \eq{Heiquad5}, the equations for the antisymmetric modes are decoupled from the pump and the solutions are, in the frequency domain \cite{Reynaud1987},
\begin{align}
\delta \tilde Q_{-i}^\mathrm{out}(\Omega)&=-\frac{i \Omega }{2 k_{a}+i \Omega}\,\delta \tilde Q_{-i} ^\mathrm{in}(\Omega)\\
\delta \tilde P_{-i}^\mathrm{out}(\Omega)&= \left(-1-\frac{2 i k_a}{\Omega}\right) \delta \tilde P_{-i}^\mathrm{in}(\Omega).
\end{align}
The frequency-domain equations for the symmetric modes are \cite{Reynaud1987}
\begin{align}
i\Omega \delta \tilde Q_{i+}(\Omega)&=4\chi\alpha_{i}\delta\tilde{Q}_{p}(\Omega)+\sqrt{2k_{a}}\delta\tilde{Q}_{i+}^\mathrm{in}(\Omega)
\label{eq:freqheiseqn1}\\
i\Omega{\delta\tilde P_{i+}}(\Omega)&=4\chi\alpha_{i}\delta\tilde{P}_{p}(\Omega)-2k_{a}\delta\tilde{P}_{i+}(\Omega)+\sqrt{2k_{a}}\delta\tilde P_{i+}^\mathrm{in}(\Omega) \\
i\Omega \delta\tilde Q_{p}(\Omega)&=-2\chi\sum_{i}^{n}\alpha_{i}\delta\tilde{Q}_{i+}(\Omega)-k_{p}\delta\tilde Q_{p}(\Omega)+\sqrt{2k_{p}}
\delta\tilde Q_p^\mathrm{in}(\Omega) 
 \\
i\Omega \delta\tilde P_{p}(\Omega)&=-2\chi\sum_{i}^{n}\alpha_{i}\delta\tilde{P}_{i+}(\Omega)-k_{p}\delta\tilde P_{p}(\Omega)+\sqrt{2k_{p}}\delta\tilde P_p^\mathrm{in}(\Omega). 
\label{eq:freqheiseqn}
\end{align}
These equations can be easily solved for pump and signal- idler pairs. The output quadratures are finally determined using input-output relations:
\begin{align}
\delta{Q_\pm^\mathrm{out}}&=\sqrt{2k_{a}} \delta{Q_\pm}-\delta{Q_\pm^\mathrm{in}}\nonumber \\
\delta{P_\pm^\mathrm{out}}&=\sqrt{2k_{a}} \delta{P_\pm}-\delta{P_\pm^\mathrm{in}} \nonumber\\
\delta{Q_p^\mathrm{out}}&=\sqrt{2k_{p}} \delta{Q_p}-\delta{Q_p^\mathrm{in}}\nonumber \\
\delta{P_p^\mathrm{out}}&=\sqrt{2k_{p}} \delta{P_p}-\delta{P_p^\mathrm{in}}. 
\end{align}
The solutions of \eqs{freqheiseqn1}{freqheiseqn} are 
\begin{align}
\delta \tilde Q_{+i}^\mathrm{out}(\Omega)&=
  -\left(1+\frac{2 k_a \left\{k_p \Omega+i [\Omega^2+8 \chi^2 (\alpha_i^2-\sum_j \alpha _j^2)]\right\}}{\Omega \left(-i k_p \Omega+\Omega^2-8 \chi^2 \sum_j \alpha _j^2\right)}\right)\delta \tilde Q_{+,i}^\mathrm{in}(\Omega) \nonumber \\
&\quad -\frac{16 i \chi^2 k_a \alpha _i }{\Omega \left(-i k_p \Omega+\Omega^2-8 \chi^2 \sum_j \alpha _j^2\right)}\sum_{j \neq i} \alpha _j \,\delta \tilde Q_{+,j}^\mathrm{in}(\Omega)\nonumber \\
&\quad -\frac{8 \chi \sqrt{k_a k_p} \alpha_i}{-i k_p \Omega+\Omega^2-8 \chi^2 \sum_j \alpha _j^2}\, \delta \tilde Q_{p}^\mathrm{in}(\Omega)\label{eq:solutiona}
\end{align}
\begin{align}
\delta \tilde P_{+i}^\mathrm{out}(\Omega)&=\left(-1+\frac{2 k_a}{2 k_a+i \Omega}-\frac{16 \chi^2 k_a \alpha _i^2}{(2 k_a+i \Omega) [(2 k_a+i \Omega) (k_p+i \Omega) 
+8 \chi^2 \sum_j \alpha _j^2]}\right)\delta \tilde P_{+,i}^\mathrm{in}(\Omega)  \nonumber \\
 &\quad -\frac{16 \chi^2 k_a \alpha _i }{(2 k_a+i \Omega) [(2 k_a+i \Omega) (k_p+i \Omega)+8 \chi^2 \sum_j \alpha _j^2]} \sum _{j \neq i}\alpha _j \,\delta \tilde P_{+,j}^\mathrm{in}(\Omega)\nonumber\\
 &\quad +\frac{8 \chi \sqrt{k_a k_p} \alpha _i}{(2 k_a+i \Omega) (k_p+i \Omega)+8 \chi^2 \sum_j \alpha _j^2}\, \delta \tilde P_{p}^\mathrm{in}(\Omega)
\end{align}
\begin{align}
\delta \tilde Q_{p}^\mathrm{out}(\Omega)&= \frac{k_p \Omega-i \left(\Omega^2-8 \chi^2 \sum_j \alpha _j^2\right)}{k_p \Omega+i \left(\Omega^2-8 \chi^2 \sum_j \alpha _j^2\right)}\,\delta \tilde Q_{p}^\mathrm{in}(\Omega)\nonumber\\
&\quad +\frac{4 i \chi \sqrt{k_ak_p}}{k_p \Omega+i \left(\Omega^2-8 k^2 \sum_j \alpha _j^2\right)} \sum_{j=1}^{n} \alpha _j\, \delta \tilde Q_{+,j}^\mathrm{in}(\Omega)
\end{align}
\begin{align}
\delta \tilde P_{p}^\mathrm{out}(\Omega)&=\frac{ 2 k_{a} (k_{p}-i \Omega)+i k_{p} \Omega+\Omega^2-8 \chi^2 \sum_j \alpha _j^2}{2 k_{a} (k_{p}+i \Omega)+i k_{p} \Omega-\Omega^2+8 k^2 \sum_j \alpha _j^2}\,\delta \tilde P_{p}^\mathrm{in}(\Omega)  \nonumber \\
&\quad -\frac{4 k \sqrt{k_{a}} \sqrt{k_{p}}}{2 k_{a} (k_{p}+i \Omega)+i k_{p} \Omega-\Omega^2+8 k^2 \sum_j \alpha _j^2}\sum_{i=1}^{n} \alpha _i \,\delta \tilde P_{+,i}^\mathrm{in}(\Omega).\label{eq:solutionb}         
\end{align}
Substituting the classical solutions \eq{thresh} in \eqs{solutiona}{solutionb} and taking $\Omega=0$, these equations yield: 
\begin{align}
\delta Q_{-i}^\mathrm{out}&\longrightarrow  0 \\
\delta P_{-i}^\mathrm{out}&\longrightarrow \infty \\
\delta Q_{+i}^\mathrm{out}&\longrightarrow \infty  \\
\delta P_{+i}^\mathrm{out}&=\frac{4 \chi \sqrt{k_a k_p}  \alpha _i}{k_a k_p \sqrt{\sigma }}\delta P_{p}^\mathrm{in}-\frac{4 \chi^2  \alpha _i^2}{k_a k_p \sqrt{\sigma }}\delta P_{+,i}^\mathrm{in}-\frac{4 \chi^2 \alpha _i }{k_a k_p \sqrt{\sigma }}\sum_{j \neq i} \alpha _j\, \delta P_{+j}^\mathrm{in}%\label{eq:sqtermsw0}
\end{align}

 \subsubsection{Two-mode squeezing}

In order to quantitatively study squeezing behavior, we assumed equal classical mean values for all pairs. However, we can assume any ratio between the classical mean values, as long as they satisfy $\sum_{i}^{n}|\alpha_{i}|^{2}=\frac{k_a k_{p}(\sqrt{\sigma}-1)}{4\chi^{2}}=\rm const$, \eq{thresh}.
Therefore the variances of squeezed and antisqueezed quadratures are
\begin{align}
V(Q_{-i})&=\av{(\delta Q_{-i})^2}\longrightarrow  0\\
V(P_{-i})&=\av{(\delta P_{-i})^2}\longrightarrow \infty \\ 
V(Q_{+i})&=\av{(\delta Q_{+i})^2}\longrightarrow \infty  \\
V(P_{+i})&=\av{(\delta P_{+i})^2}=\frac{2 (\sigma-1)}{n \sigma },
\label{eq:sqtermsw0}
\end{align}
which yields the classic EPR result at threshold ($\sigma=1$), the generation of $n$ independent entangled $(i,-i)$ pairs.  
 
 However, if the OPO is above threshold ($\sigma>1$), the variance of the phase sum, \eq{sqtermsw0} will increase from zero \cite{Reynaud1987,Reid1988} and will eventually stop being squeezed. It states the well-known fact that for the OPO operating well above the emission threshold, twin pairs are not independent EPR pairs due to the pump statistics, unless $k_p\ll k_a$ \cite{Reid1988}. This is precisely the mechanism that we rely upon to create multipartite entanglement in this work. We give two preliminary examples before turning to the evaluation of precise multimode entanglement criteria. 
 
 \subsubsection{Multimode squeezing}
 
We give two examples of squeezed multimode operators which will be useful in the next section.

First off, specifically combining phase sum operators of two pairs $i$ and $j$ yields 
\begin{eqnarray}
\alpha_i(P_j+P_{-j})-\alpha_j(P_i+P_{-i}) \longrightarrow 0.
\end{eqnarray}
Even though phase sum operators for each pair become noisier with increasing input pump, this noise  can be canceled appropriate linear combinations.

Another interesting example is that of operator $\sum_{i=1}^{n} (P_i+P_{-i})-x P_{p}$. 
%%%%%%%%%%%%%%%%%%%%%%%%%%%%%%%%%%%%%%%%%%%%%%%%%%%
\begin{figure}[htbp]
\begin{center}
\includegraphics[width=0.5\columnwidth]{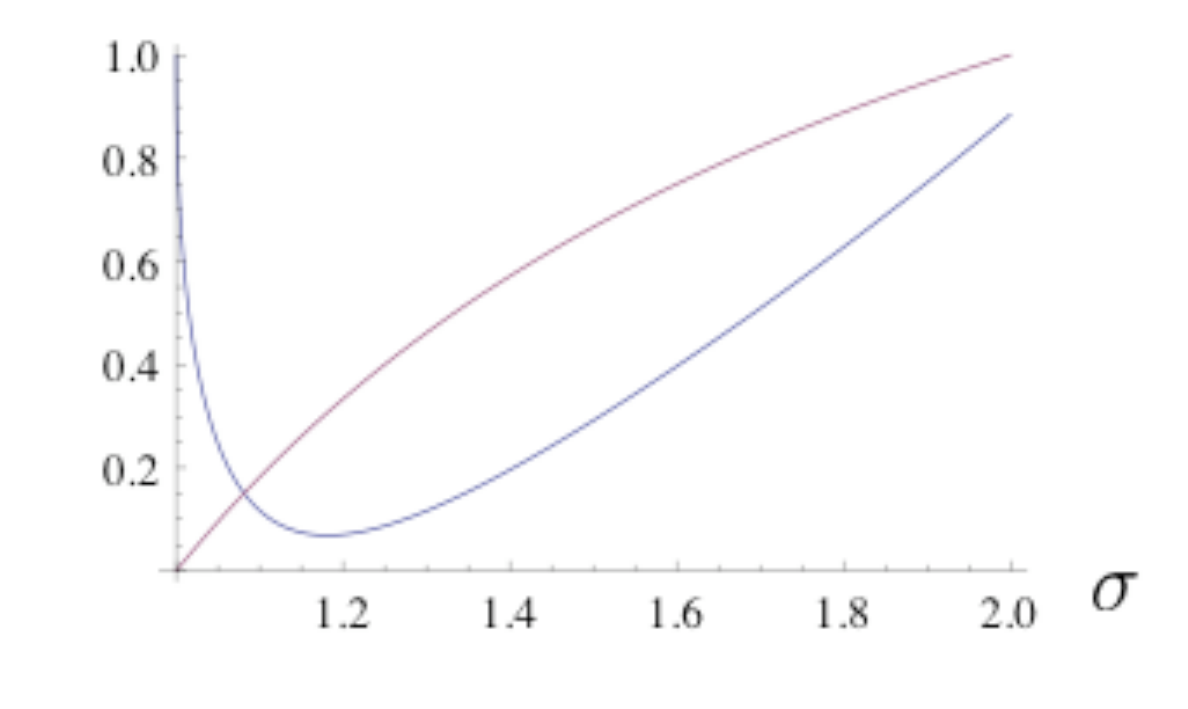}
\caption{Plot of $n V(P_i+P_{-i})$, in red, and $V[\sum_{i=1}^{n} (P_i+P_{-i})-x P_{p}]$, in blue, for $n=3$ and $x=\sigma$. When $\sigma$ increases, the variance of $P_i+P_{-i}$ increases from zero and approaches the shot noise level. However, the variance of  $\sum_{i=1}^{n} (P_i+P_{-i})-x P_{p}$, not squeezed at threshold, subsequently drops from the shot noise level (of value 1 in this graph) and shows squeezing. 
In this particular graph, the minimum of $V [\sum_{i=1}^{n} (P_i+P_{-i})-x P_{p}]$ occurs at $\sigma=1.18$ but, in general, the value of $\sigma$  for which the blue curve reaches its minimum, as well as the value of the minimum itself, depends on the choice of $x$.}
\label{fig:sq}
\end{center}
\end{figure}
%%%%%%%%%%%%%%%%%%%%%%%%%%%%%%%%%%%%%%%%%%%%%%%%%%%
In \fig{sq} we plot its variance as a function of $\sigma$, for $x=\sigma$. This graph clearly shows that the assumption of the existence of a correlation between all modes and the pump is a sensible one. Having established this, we turn to directly testing the existence of multipartite entanglement in our system.

\section{Multipartite entanglement in the OPO well above threshold}

\subsection{The van Loock-Furusawa inseparability criterion}

The van Loock-Furusawa (vLF) multipartite entanglement criterion \cite{vanLoock2003a} is the multipartite generalization of the Duan \cite{Duan2000}-Simon \cite{Simon2000} criterion, itself the continuous-variable formulation of the Peres \cite{Peres1996}-Horodecki \cite{Horodecki1996} positive partial transpose criterion. A density operator is partially separable if and only if it can be written as the convex sum
\begin{align}
\hat{\rho}=\sum_{i}\eta_{i}\hat{\rho}_{i,k_{1}.....k_{m}}\otimes \hat{\rho}_{i,k_{m+1}\dots,k_{n}},\label{eq:rho}
\end{align}
where the mode set $(k_{1},...,k_{m})$ is separable from the mode set  $(k_{m+1},\dots,k_{n})$. 
If we define two ``entanglement witnesses,'' quadrature operators with arbitrary real parameter sets $\{h_{i}\}_i$ and $\{g_{i}\}_i$, 
\begin{align}
u=h_{1}Q_{1}+h_{2}Q_{2}+\dots+h_{n}Q_{n}\\
v=g_{1}P_{1}+g_{2}P_{2}+\dots+g_{n}P_{n},
\end{align}
then the separable density operator of \eq{rho} must verify the vLF inequality \cite{vanLoock2003a}
\begin{align}
V_{\rho} (u)+V_{\rho} (v) \geqslant   
%& 
2\(|h_{k_{1}}g_{k_{1}}+\dots+h_{k_{m}}g_{k_{m}}| %\nonumber\\
%& 
+|h_{k_{m+1}}g_{k_{m+1}}+\dots+h_{k_{n}}g_{k_{n}}|\),
\label{eq:vanlookineq}
\end{align}
whose violation implies the existence of entanglement between mode set ($k_{r}$,\dots,$k_{m}$) and mode set ($k_{s}$,\dots,$k_{n}$). Operators $u$ and $v$ were coined {\em variance-based entanglement witnesses for continuous-variable systems} by Hyllus and Eisert \cite{Hyllus2006}, in reference to the original Qbit expectation-value-based entanglement witnesses \cite{Horodecki1996,Terhal2000,Lewenstein2000}.

\subsection{Multipartite entanglement in a single, depleted-pump OPO}

In order to demonstrate multipartite entanglement, we examine the conditions for violation of \emph{all} possible vLF inequalities, corresponding to all possible respective mode partitions such as \eq{rho}, and their associated experimental regimes. 

\subsubsection{Pump-signals partition}

\begin{figure}[htbp]
\begin{center}
\parbox{0.1\columnwidth}{\vfill\includegraphics[width=0.1\columnwidth]{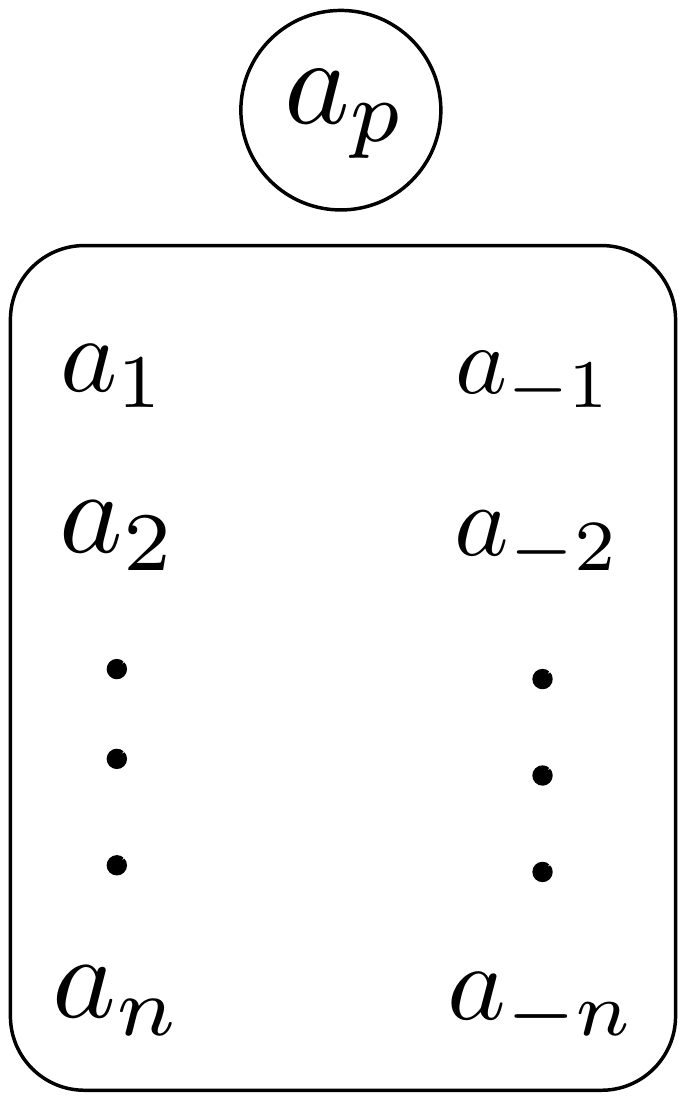}\vfill}\hfill
\parbox{0.35\columnwidth}{\includegraphics[width=0.35\columnwidth]{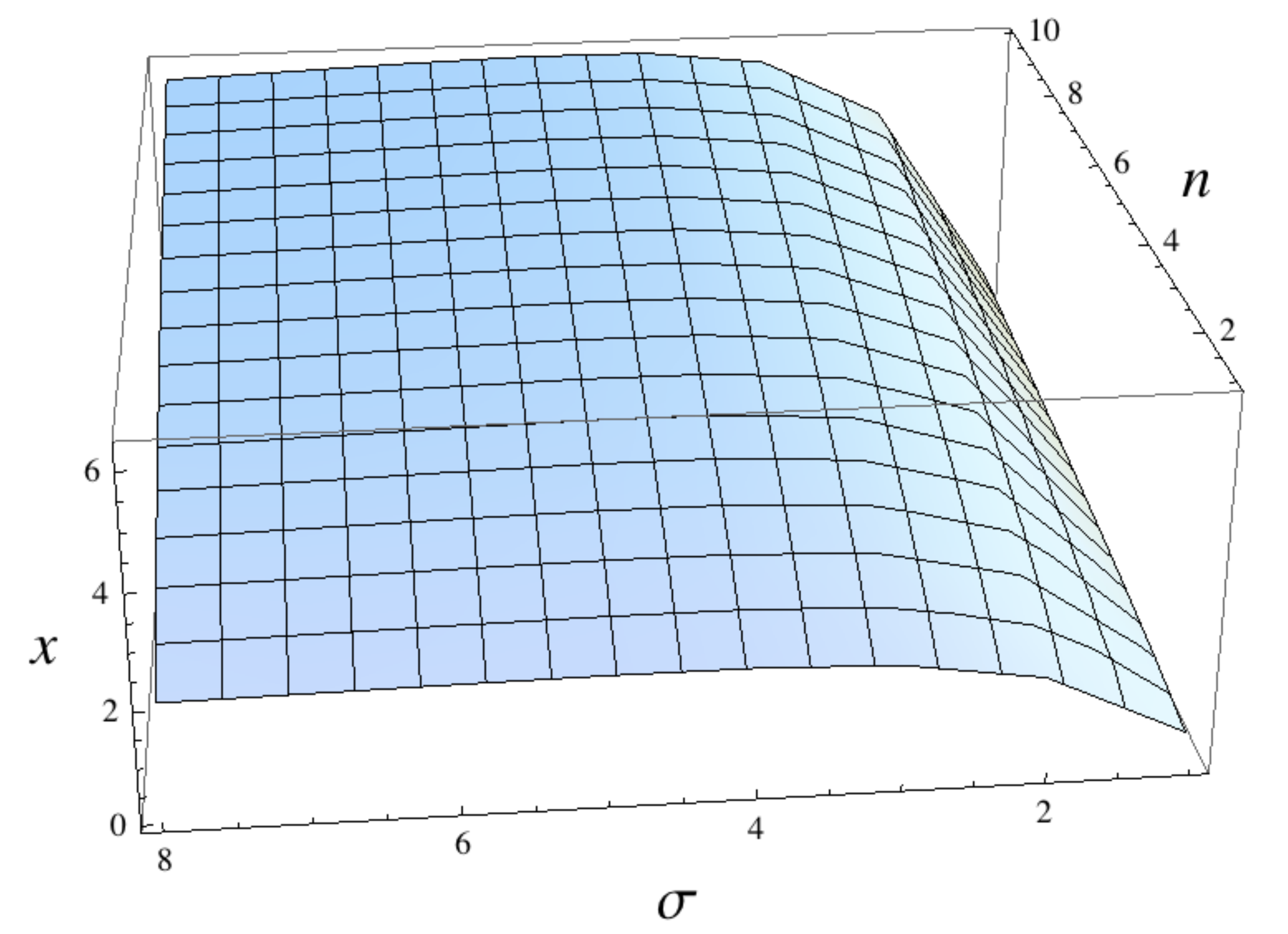}}\hfill
\parbox{0.45\columnwidth}{\includegraphics[width=0.45\columnwidth]{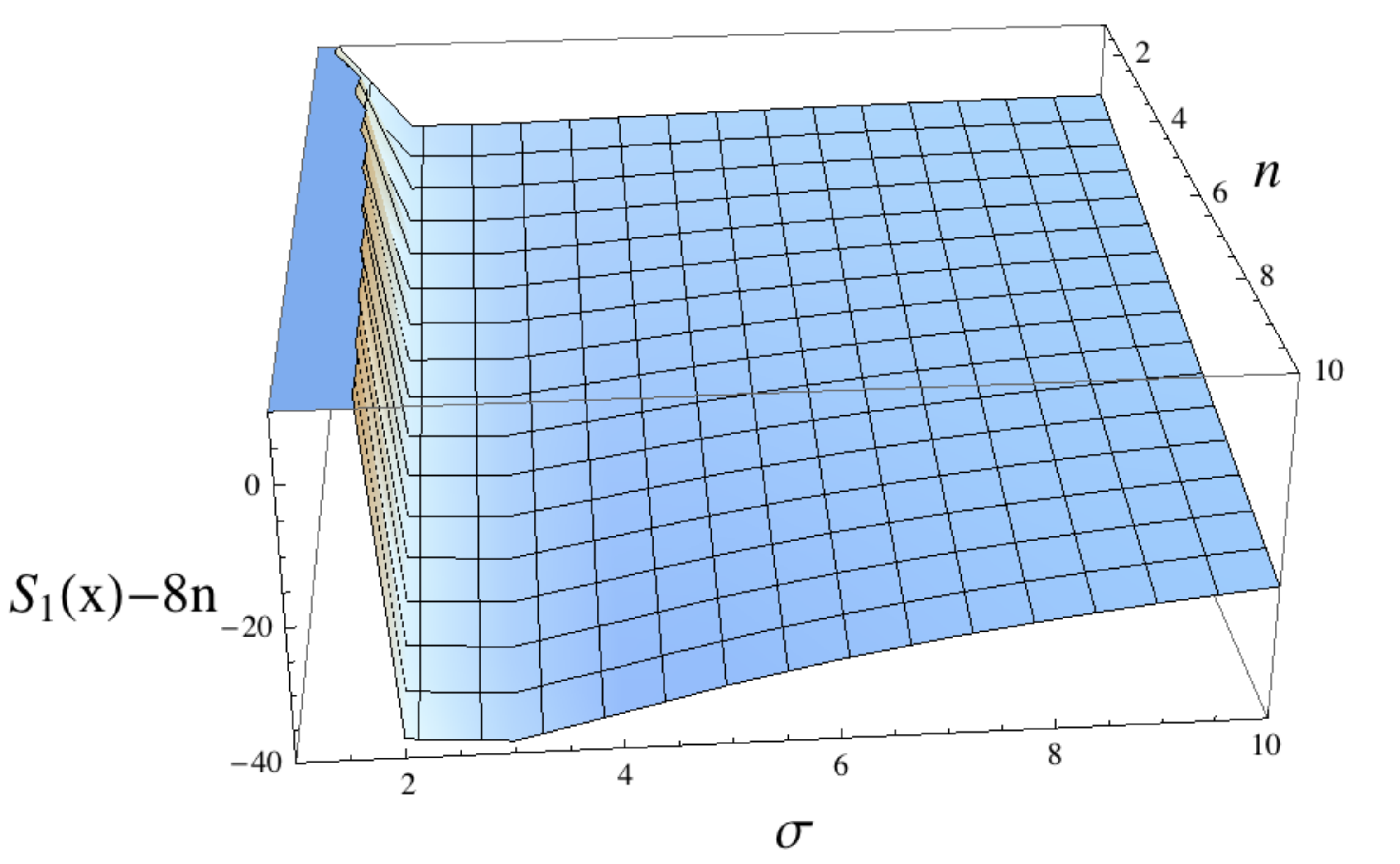}}
\caption{Left, sketch of the mode partition studied. Center, plot of the optimum values of $x=x_\mathrm{opt}$ which give maximum violation $S_{1}(x)-8n$ of the vLF inequality, at a given pump to threshold power ratio $\sigma$ and a given number $n$ of mode pairs inside the cavity. Right, plot of the maximum vLF inequality violation $S_{1}(x_\mathrm{opt})-8n$, versus $\sigma$ and $n$. We took the particular case $\Omega=0$ and $\alpha_{i}=\alpha_{j} $   $ \forall  i,j$.}
\label{fig:S1}
\end{center}
\end{figure}

We first consider the separability of the sole pump mode from all  signal modes. We define $u_{1}$ and $v_{1}$ as
\begin{align}
u_1&=\sum_{i=1}^{n}\frac{\alpha_{i}}{\alpha_{1}}(Q_{{i}}+Q_{-i})+\frac{2}{x}\sum_{i=1}^{n}\frac{\alpha_{i}}{\alpha_{1}}Q_{p}\\
v_1&=\sum_{i=1}^{n}(P_{i}+P_{-i})-x P_{p},
\end{align}
with the real parameter $x>0$. Based on \eq{vanlookineq}, the separability of mode $p$ implies    
\begin{align}
 S_{1}= V(u_1) + V(v_1) &\geqslant 2 \(|h_{-n}g_{-n}+\dots+h_{-1}g_{-1}+h_{1}g_{1}+\dots+h_{n}g_{n}|+|h_{p}g_{p}|\)
 \\
 &\geqslant 2 \(2|h_{1}g_{1}+\dots+h_{n}g_{n}|+|h_{p}g_{p}|\) \\
&\geqslant
2(|\frac{2}{\alpha_{1}}\sum_{i=1}^{n}\alpha_{i}|+|\frac{2}{x\alpha_{1}}\sum_{i=1}^{n}\alpha_{i}(-x )|)\\
&\geqslant\frac{8}{\alpha_{1}}\sum_{i=1}^{n}\alpha_{i}.
\end{align}
Here, and in the following, we make the assumption that all classical amplitudes $\{\alpha_i\}_i$ are equal, for the sake of simplicity. This doesn't lessen the generality of our treatment but makes numerical evaluations easier. Under this assumption, we get
\begin{equation}
S_1\geqslant 8n. 
\end{equation}
\Fig{S1} displays the maximum violation of the above inequality versus $n$ and $\sigma$, for optimized values of the arbitrary weight $x$. As can be seen, there always exist values of $(n,\sigma)$ for which $S_{1}-8 n$ is negative, which proves the inseparability of the pump mode from the signal modes. Unsurprisingly, entangling a larger number of pairs ($n$) requires a higher pump to threshold power ratio ($\sigma$). However, a large pump power ($\sigma$) does degrade the inseparability, which might just be due to the increasing depletion of the intracavity pump field.
%%%%%%%%%%%%%%%%%%%%%%%%%%%%%%%%%%%%%%%%%%%%%%%%%%%
%%%%%%%%%%%%%%%%%%%%%%%%%%%%%%%%%%%%%%%%%%%%%%%%%%%

\subsubsection{Partition of one $(a_{j},a_{-j})$ EPR pair}

\begin{figure}[htbp]
\begin{center}
\parbox{0.1\columnwidth}{\vfill\includegraphics[width=0.1\columnwidth]{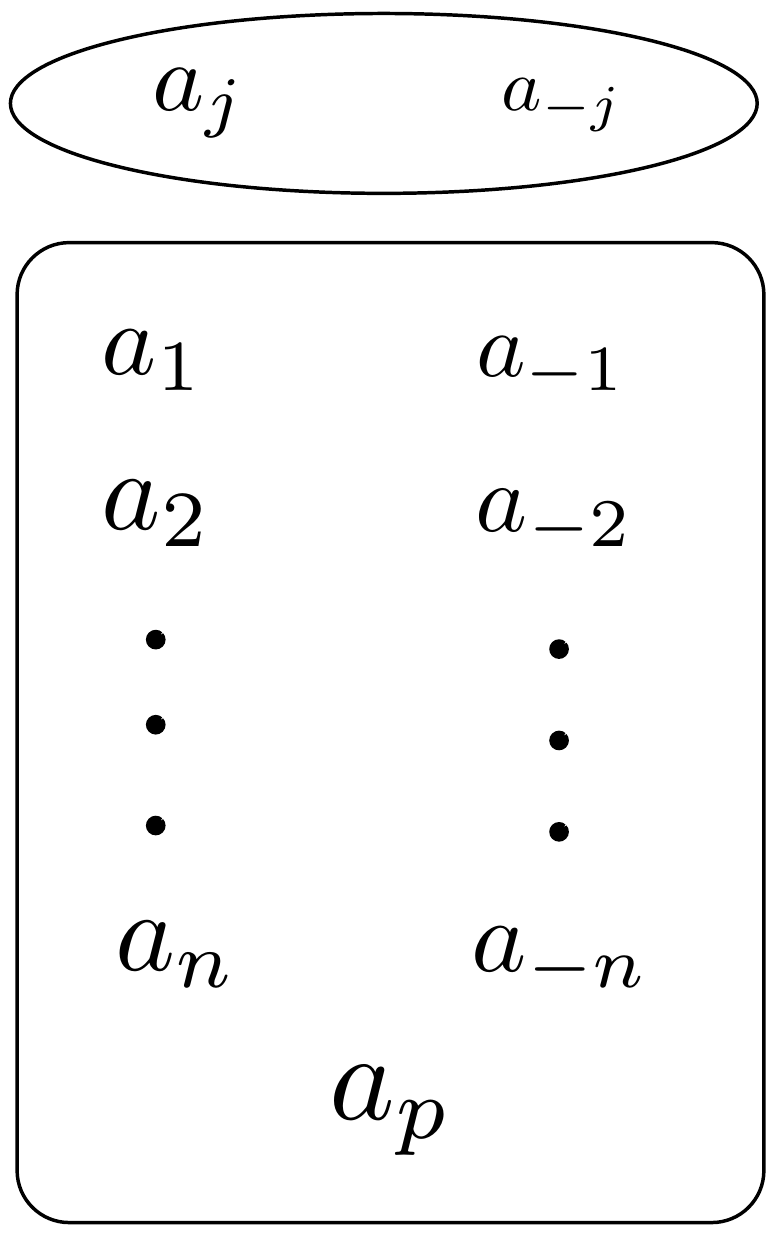}\vfill}\hfill
\parbox{0.35\columnwidth}{\includegraphics[width=0.35\columnwidth]{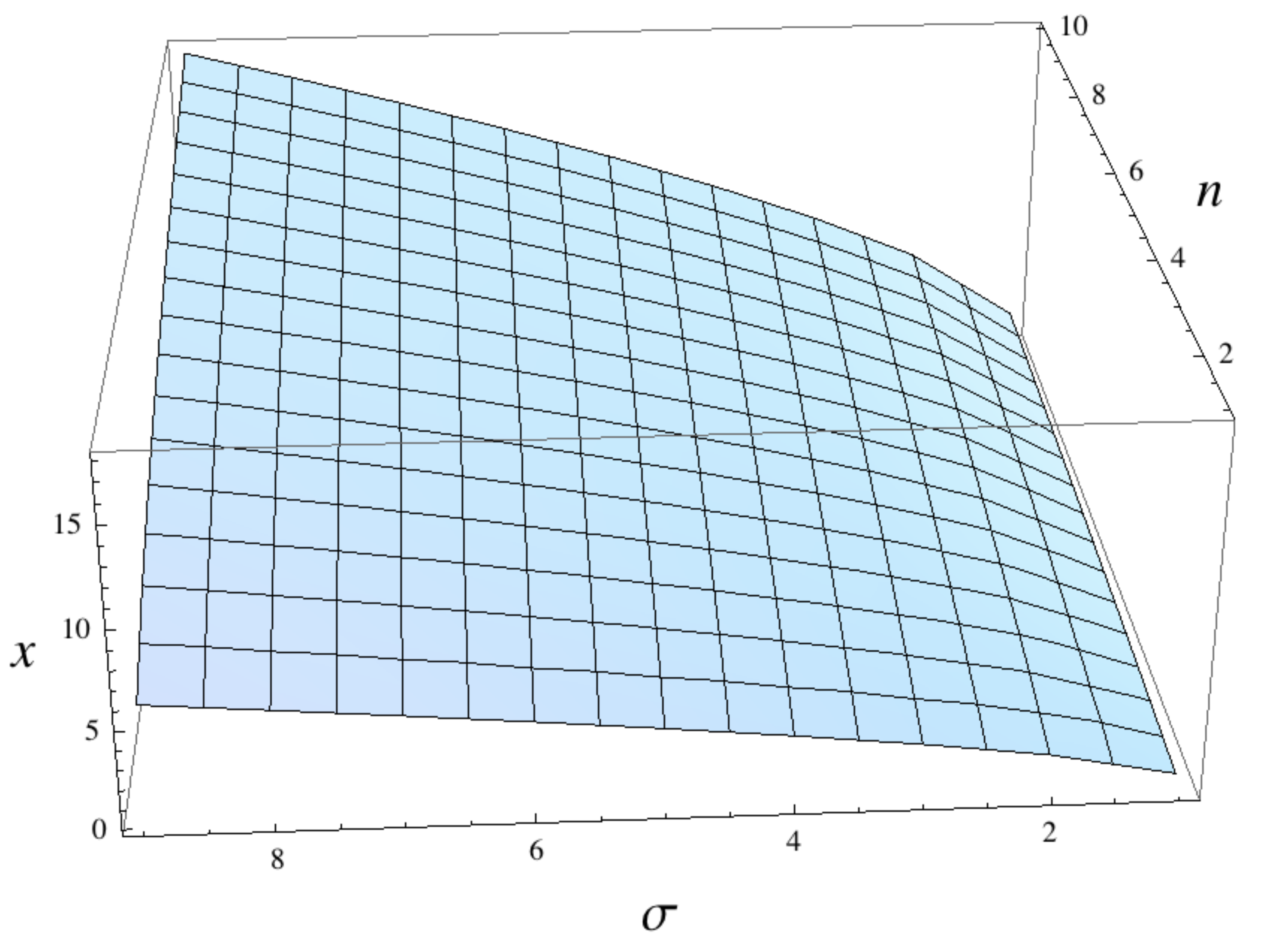}}\hfill
\parbox{0.45\columnwidth}{\includegraphics[width=0.45\columnwidth]{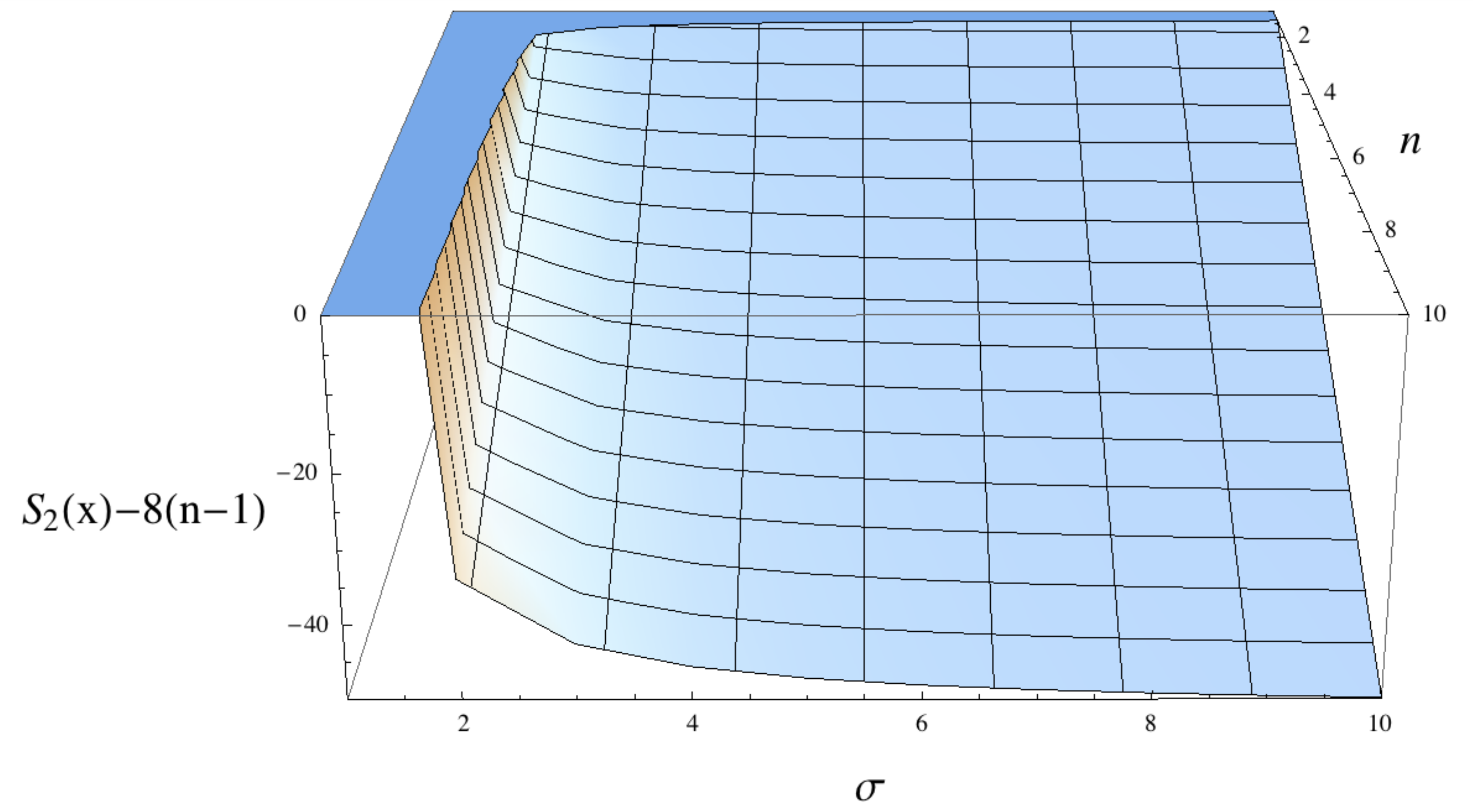}}
\caption{Left, sketch of the mode partition studied. Center, plot of the optimum values of $x=x_\mathrm{opt}$ which give maximum violation $S_{2}(x)-8(n-1)$ of the vLF inequality, at a given pump to threshold power ratio $\sigma$ and a given number $n$ of mode pairs inside the cavity. Right, plot of the maximum vLF inequality violation $S_{2}(x_\mathrm{opt})-8(n-1)$, versus $\sigma$ and $n$. We took the particular case $\Omega=0$ and $\alpha_{i}=\alpha_{j} $   $ \forall  i,j$.}
\label{fig:S2}
\end{center}
\end{figure}

We now study the inseparability of an entangled pair $(a_{j},a_{-j})$ from the rest of the signals and the pump. For such a partition, we define
\begin{align}
u_2&=Q_{j+}+\sum_{i\neq j}^{n}\frac{\alpha_{i}}{\alpha_{j}}Q_{i+}+\frac{2}{x\alpha_{j}}\sum_{i}^{n}\alpha_{i}Q_{p}\\
v_{2}&=\sum_{i\neq j}^{n}(\frac{\alpha_{i}}{\alpha_{j}}P_{j+}-P_{i+}),
\end{align}
and the vLF inequality is
\begin{align}
S_{2} = V(u_2) +V(v_2) &\geqslant 
 2 \( |2h_{j}g_{j}|+|2h_{1}g_{1}+\dots+2h_{n}g_{n}+h_{p}g_{p}|\) \\
& \geqslant 2\left|\(2\frac{1}{\alpha_{j}}\sum_{i\neq j}^{n}\alpha_{i}\)\right|+2\left|-\frac{2}{\alpha_{j}}\sum_{i\neq}^{n}\alpha_{i}+0\right| \\
& \geqslant
 \frac{8}{\alpha_{j}}\sum_{i\neq j}^{n}\alpha_{i} \\
 & \geqslant  8(n-1).
 \label{eq:S2}
\end{align}
\Fig{S2} shows the violation of this inequality for a broad range of parameters. It also demonstrates the necessity of applying larger input pump intensity when considering more pairs inside the cavity in order to generate inseparability  between all pairs, again unsurprisingly. 
%%%%%%%%%%%%%%%%%%%%%%%%%%%%%%%%%%%%%%%%%%%%%%%%%%% 
%%%%%%%%%%%%%%%%%%%%%%%%%%%%%%%%%%%%%%%%%%%%%%%%%%%
 
\subsubsection{Pair set partition}\label{sec:EPR}

We next turn to  partitions $\{(a_{1},b_{1})\dots(a_{k},b_{k})\}\{(a_{k+1},b_{k+1})\dots(a_{n},b_{n})\}$. The operators are
\begin{align}
u_3&=\sum_{i=1}^{k}Q_{i+}
+\sum_{j=k+1}^{n}\frac{\alpha_{j}}{\alpha_{1}}Q_{j+}+\frac{2}{x\alpha_{1}}\sum_{l}^{n}\alpha_{l}Q_{p}\\
v_3&=\sum_{i=1}^{k}\sum_{j\neq i}^{n}(P_{i+}-\frac{\alpha_{i}}{\alpha_{j}}P_{j+}),
\end{align}
and the vLF inequality is
\begin{align}
S_{3} = V(u_3) + V(v_3) 
&\geqslant  
 2\(|2h_{1}g_{1}+\dots+2h_{k}g_{k}| +|2h_{k+1}g_{k+1}+\dots+2h_{n}g_{n}+h_{p}g_{p}|\)  \label{eq:s3} \\
& \geqslant 8k(n-k).
\end{align}
Assuming $1\leqslant k < n$, then it is straightforward to show that $8(n-1)\leqslant 8k(n-k) \leqslant 2n^{2}$. As a consequence, the vLF inequality $S_{3}$ is automatically violated when vLF inequality $S_{2}$ is, and does not need to be considered separately. 

\subsubsection{Single signal mode partition}\label{sec:nonEPR}

Finally, we consider partitions of a single signal mode, or several such, all belonging to different EPR pairs $(i,-i)$.
Because the signal mode $a_{i}$ is highly entangled to the mode  $a_{-i}$---and to all other signals by virtue of the preceding---the separability test is simple in the present case. It is straightforward to show that the necessary vLF inequality for the partition $\{a_{1},\dots,a_{k}\}\{a_{-1},\dots,a_{-k},
(a_{k+1},a_{-(k+1)}),\dots,(a_n,a_{-n}))\}$ ,

 \begin{align}
S_{4}=&V\left(Q_{j}-Q_{-j}
+\sum_{i\neq j}\frac{\alpha_{i}}{\alpha_{j}}(Q_{i}-Q_{-i})\right) 
+V\left(\sum_{i\neq j}P_{i}+P_{-i}-\frac{\alpha_{i}}{\alpha_{j}}(P_{j}+P_{-j})\right) \nonumber\\
&\geqslant 2 \( |h_{1}g_{1}+\dots+h_k g_{k}|+|h_{-1}g_{-1}+\dots h_{-k}g_{-k}+2h_{k+1}g_{k+1}+\dots 2h_{n}g_{n}|\)\nonumber\\
  &\geqslant 2|-\sum_{i\neq1} \frac{\alpha_i}{\alpha_1}+\sum_{i=2}^{k} \frac{\alpha_k}{\alpha_1}|+2|\sum_{i\neq1} \frac{\alpha_{-i}}{\alpha_1}+\sum_{i=2}^{k} \frac{\alpha_{-k}}{\alpha_1}+0|\nonumber\\
 &\geqslant 4(n-k),
\label{eq:s32}
\end{align}
is always violated in the presence of single EPR pair entanglement. This is because the left hand side term of \eq{s32} contains EPR nullifiers \cite{Gu2009}, a.k.a.\ EPR entanglement witnesses, whose squeezed variances tend toward zero. 
The inseparability of any other form of partitions on modes when modes $a_{1}$ and $a_{-1}$ are placed in different partitions, can be examined by inequalities similar to $S_4$ and with nonzero boundaries. Such inequalities are always violated.  
Therefore, if EPR entanglement is present (the checking of which is a staple of the experimental calibration of a regular two-mode squeezer), then the violation of both $S_{1}$ and $S_{2}$ is a  necessary and sufficient condition to mode inseparability for all possible partitions in the optical frequency comb of a single OPO.

\subsection{Entanglement between pairs without considering the pump}

Here we ask the question of the possibility of multipartite entanglement between twins without considering the pump field. For that, we rewrite inequality $S_{2}$ without the pump quadratures: 
\begin{align}
S_{2}^{'}= V(u_2^{'})+ V(v_2^{'}) \geqslant 8(n-1)
\end{align}
with
\begin{align}
u_2^{'}=Q_{j+}+\sum_{i \neq j}^{n}\frac{\alpha_{i}}{\alpha_{j}}Q_{i+} \\
v_2^{'}=\sum_{i\neq j}^{n}\frac{\alpha_{i}}{\alpha_{j}}P_{j+}-P_{i+}.
\end{align}
%%%%%%%%%%%%%%%%%%%%%%%%%%%%%%%%%%%%%%%%%%%%%%%%%%% 
\begin{figure}[htbp]
\begin{center}
\includegraphics[width=0.5\columnwidth]{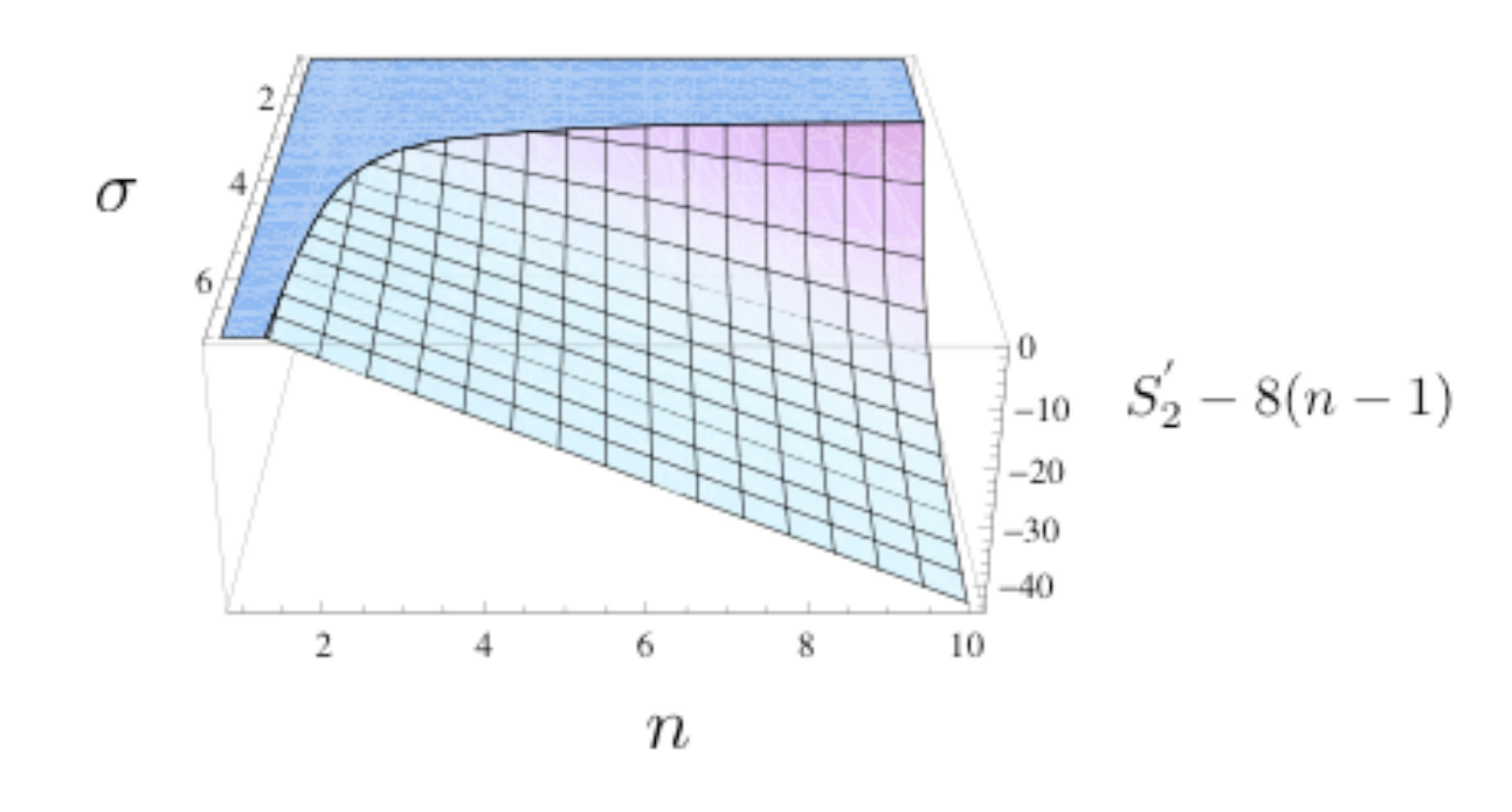}
\caption{Plot of the vLF inequality violation $S_{2}^{'}-8(n-1)$, versus $\sigma$, the pump to threshold power ratio, and $n$, the number of mode pairs inside the cavity, for $\Omega=0$ and $\alpha_{i}=\alpha_{j}$,  $\forall  i,j$.}
\label{fig:S2prime}
\end{center}
\end{figure}
%%%%%%%%%%%%%%%%%%%%%%%%%%%%%%%%%%%%%%%%%%%%%%%%%%%
In \fig{S2prime}, we plot the violation of this inequality. As can be seen by comparing with \fig{S2}, Inequality $S_{2}^{'}$ requires slightly larger $\sigma$ to be violated, compared to $S_{2}$,  for small values of n. It shows we need to pump harder (and get closer to total depletion) in order to see pure entanglement between twin pairs. %However, as seen from \fig{S2prime},  $S_{2}$ and $S_{2}^{'}$ are violated for similar values of $\sigma$ when $n > 6$. 
The arguments of subsections and \ref{sec:EPR} and \ref{sec:nonEPR} may be reused here to complete the inseparability proof.

\subsection{Optimal entanglement witnesses}

A valid question is whether the  entanglement witnesses that were derived here on the basis of physical considerations, namely the pump-depletion-induced correlations between EPR pairs, are in fact the optimal entanglement witnesses for the system. In other words, do different observables exist that would lead to even more strongly violated vLF inequalities? Answering this question should, in turn, inform on what type of entangled state is really generated here. The search for optimal entanglement witnesses was addressed by Hyllus and Eisert, using semidefinite programming procedures \cite{Hyllus2006}. While the scope of the present paper is limited to this successful demonstration of large-scale entanglement in a simple OPO, the exact nature of the quantum state generated is clearly an interesting followup question, on which light can be shed by seeking the optimal entanglement witnesses and checking whether they are different from the ones derived above.

\section{Conclusion}

We showed that a single OPO operating well above threshold can generate multipartite entanglement in its quantum optical frequency comb. We verified the multipartite nature of the entanglement by evaluating the van Loock-Furusawa separability criterion over all possible Qmode partitions. We showed that all of these vLF inequalities can be violated, for an arbitrary large number of pairs $n$, simply by increasing the input pump power higher above threshold. 

While the presence of multipartite entanglement in such a simple system is a remarkable feature, it is important to keep in mind that the exact \emph{type} of entanglement that is produced here (GHZ, W, cluster) is difficult to determine. The search for optimal entanglement witnesses for this system is a promising approach to illuminate this question. Note that previous work has shown that multipartite \emph{cluster-state} generation, which was experimentally demonstrated in a single OPO below threshold \cite{Pysher2011}, should actually fail above threshold \cite{Midgley2010a}. However, the multipartite entanglement that we discovered in the simple OPO above threshold would certainly be useful for quantum communication applications, such as quantum secret sharing.

This work was supported by U.S. National Science Foundation grants No.\ PHY-0855632 and No.\ PHY-0960047.

\bibliographystyle{/Users/opfisterSL/Documents/Werk/Articles/Authors/P/Pfister/BibStyles/bibstyleNCM}

\begin{thebibliography}{10}

\bibitem{Yonezawa2004}
H. Yonezawa, T. Aoki, and A. Furusawa, ``Demonstration of a quantum
  teleportation network for continuous variables,'' Nature {\bf 431}, 430
  (2004).

\bibitem{Lance2004}
A.~M. Lance, T. Symul, W.~P. Bowen, B.~C. Sanders, and P.~K. Lam, ``Tripartite
  quantum state sharing,'' Phys.\ Rev.\ Lett. {\bf 92}, 177903 (2004).

\bibitem{Briegel2001}
H.~J. Briegel and R. Raussendorf, ``Persistent entanglement in arrays of
  interacting particles,'' Phys.\ Rev.\ Lett. {\bf 86}, 910 (2001).

\bibitem{Zhang2006}
J. Zhang and S.~L. Braunstein, ``Continuous-variable {G}aussian analog of
  cluster states,'' Phys.\ Rev.\ A {\bf 73}, 032318 (2006).

\bibitem{Raussendorf2001}
R. Raussendorf and H.~J. Briegel, ``A one-way quantum computer,'' Phys.\ Rev.\
  Lett. {\bf 86}, 5188 (2001).

\bibitem{Menicucci2006}
N.~C. Menicucci, P. {van Loock}, M. Gu, C. Weedbrook, T.~C. Ralph, and M.~A.
  Nielsen, ``Universal quantum computation with continuous-variable cluster
  states,'' Phys.\ Rev.\ Lett. {\bf 97}, 110501 (2006).

\bibitem{Mermin2003}
N.~D. Mermin, ``From {C}bits to {Q}bits: Teaching computer scientists quantum
  mechanics,'' American Journal of Physics {\bf 71}, 23 (2003).

\bibitem{Ladd2010}
T.~D. Ladd, F. Jelezko, R. Laflamme, Y. Nakamura, C. Monroe, and J.~L. O'Brien,
  ``Quantum computers,'' Nature {\bf 464}, 45 (2010).

\bibitem{Greiner2002}
M. Greiner, O. Mandel, T. Esslinger, T.~W. {H\"ansch}, and I. Bloch, ``Quantum
  phase transition from a superfluid to a Mott insulator in a gas of ultracold
  atoms,'' Nature {\bf 415}, 39 (2002).

\bibitem{Menicucci2008}
N.~C. Menicucci, S.~T. Flammia, and O. Pfister, ``One-way quantum computing in
  the optical frequency comb,'' Phys.\ Rev.\ Lett. {\bf 101}, 130501 (2008).

\bibitem{Flammia2009}
S.~T. Flammia, N.~C. Menicucci, and O. Pfister, ``The optical frequency comb as
  a one-way quantum computer,'' J. Phys.\ B, {\bf 42}, 114009 (2009).

\bibitem{Pysher2011}
M. Pysher, Y. Miwa, R. Shahrokhshahi, R. Bloomer, and O. Pfister, ``Parallel
  generation of quadripartite cluster entanglement in the optical frequency
  comb,'' Phys.\ Rev.\ Lett. {\bf 107}, 030505 (2011).

\bibitem{Midgley2010a}
S.~L.~W. Midgley, M.~K. Olsen, A.~S. Bradley, and O. Pfister, ``Analysis of a
  continuous-variable quadripartite cluster state from a single optical
  parametric oscillator,'' Phys.\ Rev.\ A {\bf 82}, 053826 (2010).

\bibitem{Pooser2005}
R.~C. Pooser and O. Pfister, ``Observation of triply coincident nonlinearities
  in periodically poled {$\rm KTiOPO_4$},'' Opt.\ Lett. {\bf 30}, 2635 (2005).

\bibitem{Pysher2010}
M. Pysher, A. Bahabad, P. Peng, A. Arie, and O. Pfister, ``Quasi-phase-matched
  concurrent nonlinearities in periodically poled {$\rm KTiPO_4$} for quantum
  computing over the optical frequency comb,'' Opt.\ Lett. {\bf 35}, 565
  (2010).

\bibitem{Vyas1995}
R. Vyas and S. Singh, ``Exact Quantum Distribution for Parametric
  Oscillators,'' Phys.\ Rev.\ Lett. {\bf 74}, 2208 (1995).

\bibitem{Dechoum2004}
K. Dechoum, P. Drummond, S. Chaturvedi, and M. Reid, ``Critical quantum
  fluctuations in the nondegenerate parametric oscillator,'' Phys.\ Rev.\ A
  {\bf 70}, 053807 (2004).

\bibitem{Bradley2005}
A.~S. Bradley, M.~K. Olsen, O. Pfister, and R.~C. Pooser, ``Bright tripartite
  entanglement in triply concurrent parametric down conversion,'' Phys.\ Rev.\
  A {\bf 72}, 053805 (2005).

\bibitem{Midgley2010}
S.~L.~W. Midgley, A.~S. Bradley, O. Pfister, and M.~K. Olsen, ``Quadripartite
  continuous-variable entanglement via quadruply concurrent down-conversion,''
  Phys.\ Rev.\ A {\bf 81}, 063834 (2010).

\bibitem{Navarrete2008}
C. Navarrete-Benlloch, E. Rold\'an, and G.~J. de~Valc\'arcel, ``Noncritically
  Squeezed Light via Spontaneous Rotational Symmetry Breaking,'' Phys.\ Rev.\
  Lett. {\bf 100}, 203601 (2008).

\bibitem{Navarrete2010}
C. Navarrete-Benlloch, A. Romanelli, E. Rold\'an, and G.~J. de~Valc\'arcel,
  ``Noncritical quadrature squeezing in two-transverse-mode optical parametric
  oscillators,'' Phys.\ Rev.\ A {\bf 81}, 043829 (2010).

\bibitem{Navarrete2009}
C. Navarrete-Benlloch, G.~J. de~Valc\'arcel, and E. Rold\'an, ``Generating
  highly squeezed hybrid Laguerre-Gauss modes in large-Fresnel-number
  degenerate optical parametric oscillators,'' Phys.\ Rev.\ A {\bf 79}, 043820
  (2009).

\bibitem{Chalopin2010}
B. Chalopin, F. Scazza, C. Fabre, and N. Treps, ``Multimode nonclassical light
  generation through the optical-parametric-oscillator threshold,'' Phys.\
  Rev.\ A {\bf 81}, 061804 (2010).

\bibitem{Reid1989a}
M.~D. Reid and P.~D. Drummond, ``Correlations in nondegenerate parametric
  oscillation: Squeezing in the presence of phase diffusion,'' Phys.\ Rev.\ A
  {\bf 40}, 4493 (1989).

\bibitem{Courtois1991}
J.~Y. Courtois, A. Smith, C. Fabre, and S. Reynaud, ``Phase diffusion and
  quantum noise in the optical parametric oscillator: a semiclassical
  approach,'' J. Mod.\ Opt. {\bf 38}, 177 (1991).

\bibitem{vanLoock2003a}
P. van Loock and A. Furusawa, ``Detecting genuine multipartite
  continuous-variable entanglement,'' Phys.\ Rev.\ A {\bf 67}, 052315 (2003).

\bibitem{Reid1989}
M. Reid, ``Demonstration of the {Einstein}-{Podolsky}-{Rosen} paradox using
  nondegenerate parametric amplification,'' Phys.\ Rev.\ A {\bf 40}, 913
  (1989).

\bibitem{Ou1992}
Z.~Y. Ou, S.~F. Pereira, H.~J. Kimble, and K.~C. Peng, ``Realization of the
  {Einstein}-{Podolsky}-{Rosen} paradox for continuous variables,'' Phys.\
  Rev.\ Lett. {\bf 68}, 3663 (1992).

\bibitem{Reid1988}
M. Reid and P. Drummond, ``Quantum correlations of phase in nondegenerate
  parametric oscillation,'' Phys.\ Rev.\ Lett. {\bf 60}, 2731 (1988).

\bibitem{Villar2005}
A. Villar, L.~S. Cruz, K.~N. Cassemiro, M. Martinelli, and P. Nussenzveig,
  ``Generation of Bright Two-Color Continuous Variable Entanglement,'' Phys.\
  Rev.\ Lett. {\bf 95}, 243603 (2005).

\bibitem{Su2006}
X. Su, A. Tan, X. Jia, Q. Pan, C. Xie, and K. Peng, ``Experimental
  demonstration of quantum entanglement between frequency-nondegenerate optical
  twin beams,'' Opt.\ Lett. {\bf 31}, 1133 (2006).

\bibitem{Jing2006}
J. Jing, S. Feng, R. Bloomer, and O. Pfister, ``Experimental
  continuous-variable entanglement from a phase-difference-locked optical
  parametric oscillator,'' Phys.\ Rev.\ A {\bf 74}, 041804(R) (2006).

\bibitem{Keller2008}
G. Keller, V. {D'Auria}, N. Treps, T. Coudreau, J. Laurat, and C. Fabre,
  ``Experimental demonstration of frequency-degenerate bright {EPR} beams with
  a self-phase-locked {OPO},'' Opt.\ Exp {\bf 16}, 9351 (2008).

\bibitem{Note1}
Note that all previous works featured the entanglement of a single Qmode pair
  at a time, which is not the situation described by Eq.~(\ref {eq:hundep}).
  Indeed, Eq.~(\ref {eq:hundep}) predicts many independent EPR pairs.
  Experimentally, this requires that the OPO cavity be resonant for all Qmode
  EPR pairs, which can be realized either by compensating birefringence in a
  type-II OPO or by using a type-I OPO. Dispersion is neglected in this
  discussion as its effects can be neglected for the first tens to hundreds of
  modes.

\bibitem{Villar2006}
A.~S. Villar, M. Martinelli, C. Fabre, and P. Nussenzveig, ``Direct Production
  of Tripartite Pump-Signal-Idler Entanglement in the Above-Threshold Optical
  Parametric Oscillator,'' Phys.\ Rev.\ Lett. {\bf 97}, 140504 (2006).

\bibitem{Coelho2009}
A.~S. Coelho, F.~A.~S. Barbosa, K.~N. Cassemiro, A.~S. Villar, M. Martinelli,
  and P. Nussenzveig, ``{Three-Color Entanglement},'' Science {\bf 326}, 823
  (2009).

\bibitem{Cassemiro2008}
K.~N. Cassemiro and A.~S. Villar, ``Scalable continuous-variable entanglement
  of light beams produced by optical parametric oscillators,'' Phys.\ Rev.\ A
  {\bf 77}, 022311 (2008).

\bibitem{Pfister2004}
O. Pfister, S. Feng, G. Jennings, R. Pooser, and D. Xie, ``Multipartite
  continuous-variable entanglement from concurrent nonlinearities,'' Phys. Rev.
  A {\bf 70}, 020302 (2004).

\bibitem{Collett1984}
M.~J. Collett and C.~W. Gardiner, ``Squeezing of intracavity and traveling-wave
  light fields produced in parametric amplification,'' Phys.\ Rev.\ A {\bf 30},
  1386 (1984).

\bibitem{Gardiner2004}
C. Gardiner and P. Zoller, {\em Quantum Noise, A Handbook of Markovian and
  Non-Markovian Quantum Stochastic Methods with Applications to Quantum
  Optics}, {\em Springer Series in Synergetics}, 3rd ed. (Springer, 2004).

\bibitem{Reynaud1987}
S. Reynaud, C. Fabre, and E. Giacobino, ``Quantum fluctuations in a two-mode
  parametric oscillator,'' J. Opt.\ Soc.\ Am.\ B {\bf 4}, 1520 (1987).

\bibitem{Duan2000}
L.-M. Duan, G. Giedke, J. Cirac, and P. Zoller, ``Inseparability Criterion for
  Continuous Variable Systems,'' Phys.\ Rev.\ Lett. {\bf 84}, 2722 (2000).

\bibitem{Simon2000}
R. Simon, ``{P}eres-{H}orodecki separability criterion for continuous variable
  systems,'' Phys.\ Rev.\ Lett. {\bf 84}, 2726 (2000).

\bibitem{Peres1996}
A. Peres, ``Separability Criterion for Density Matrices,'' Phys. Rev. Lett.
  {\bf 77}, 1413 (1996).

\bibitem{Horodecki1996}
M. Horodecki, P. Horodecki, and R. Horodecki, ``Separability of mixed states:
  necessary and sufficient conditions,'' Phys.\ Lett.\ A {\bf 223}, 1 (1996).

\bibitem{Hyllus2006}
P. Hyllus and J. Eisert, ``Optimal entanglement witnesses for
  continuous-variable systems,'' New J. Phys. {\bf 8}, 51 (2006).

\bibitem{Terhal2000}
B. Terhal, ``Bell Inequalities and the Separability Criterion,'' Phys.\ Lett.\
  A {\bf 271}, 319 (2000).

\bibitem{Lewenstein2000}
M. Lewenstein, B. Kraus, J.~I. Cirac, and P. Horodecki, ``Optimization of
  entanglement witnesses,'' Phys.\ Rev.\ A {\bf 62}, 052310 (2000).

\bibitem{Gu2009}
M. Gu, C. Weedbrook, N.~C. Menicucci, T.~C. Ralph, and P. van Loock, ``Quantum
  computing with continuous-variable clusters,'' Phys.\ Rev.\ A {\bf 79},
  062318 (2009).

\end{thebibliography}

\end{document}